\documentclass[reprint,amsmath,amssymb, aps,]{revtex4-1}
\usepackage{graphicx}
\usepackage{dcolumn}
\usepackage{bm}
\usepackage{hyperref}

\begin{document}
\title{Charge fluctuations of a Neutral  Black Hole}
\author{Marcelo Schiffer}
\affiliation{Department of Physics, Ariel University, Ariel 44837, Israel.}

\date{\today}

\begin{abstract}

In this paper we consider  charge fluctuations of a Schwarzschild black hole of mass $M$ in thermal equilibrium with a plasma consisting of photons, electrons and positrons confined inside a cavity or radius $R>>M$. Important, we do not set the black hole charge $Q$   identically  to zero at the outset; we allow it to have a small value (in the sense that $Q/M<<1$) and  then study  the  conditions for thermodynamical equilibrium; only then that we take the $Q \rightarrow 0$ limit. The density of states becomes extremely large near the horizon and a cut-off parameter $\Delta$ is introduced to prevent divergences in the field sector and  only the divergent terms in $\Delta$ are kept. Accordingly, the relevant fluctuations arise within a thin shell around the black hole. The final result is astonishingly simple:  charge fluctuations  have two independent contributions: one that comes from the black hole itself, with no respect to the plasma, and another one which arises from the plasma; both are Universal, do not depend on the black hole mass and none of them depends on the charge of the charge carriers. They differ by a numerical factor of order one if we assume the cut-off parameter to be Planckian. Therefore, we regard them  as  physically identical .  Interesting enough, the black hole's charge (squared) fluctuations are of the order $\hbar$ which is not too far from the fine structure constant. The physical origin of charge fluctuations remains raises a conundrum. Since it does not depend upon the charge of the charge carriers, it cannot be  regarded as resulting from the trade of charges between the black hole and the quantum fields and  is truly  of quantum gravity nature (albeit in a semi-classical approximation). There are at least two possible ways of interpreting our results: either  charge is trapped inside the black hole and the average must be regarded over an ensemble of black-holes or eventuality charge fluctuations arise from particles crossing the Cauchy horizon $\overline{r_-} \sim \hbar/2M$ in both directions. 
 \begin{description}
\item[PACS numbers : 04.70Dy,  05.40.-a,  05.70-a,52.25 Kn]
\end{description}
\end{abstract}
\maketitle
\begin{widetext}
\section{A static black hole in equilibrium with radiation}
A  black hole of mass $M$ and charge  $Q$ is described by  Reissner–Nordström metric:
\begin{equation}
ds^2=-g(r) dt^2 +g^{-1}(r) dr^2 +r^2 (d\theta^2+\sin^2\theta d\varphi^2)\, ,
\end{equation}
where 
\begin{equation}
g(r)=\frac{(r-r_+)(r-r_- )}{r^2}\, ,
\end{equation}
with
\begin{equation}
r_\pm=M\pm\sqrt{M^2-Q^2}\, .
\end{equation}
The event horizon is located at $r=r_+$ and the corresponding  area is
\begin{equation}
A=4\pi r_+^2\, .
\end{equation}

The vector potential and the electric field are given by
\begin{eqnarray}
A=(\frac{Q}{r},0,0,0) \, ,\\ 
E=(0,\frac{Q}{r^2},0,0)\, .
\end{eqnarray}

Now, we consider such a black hole in thermal equilibrium with a plasma of electrons, positrons and photons confined inside a cavity of radius $R$.  The total energy and charge of the system are:
\begin{eqnarray}
{\cal E}&=&M+E_+ + E_- +E_\gamma  \, ,\nonumber \\
{\cal Q}&= &Q+e(N_+-N_-) \, ,
\label{conservation}
\end{eqnarray}
where 
\begin{eqnarray}
E_\pm&=&\sum_s \int \frac{n_\pm(\epsilon) \epsilon  }{e^{\beta (\epsilon \mp \mu )}+1}d\epsilon  \, ,\\
E_\gamma&=&\sum_s  \int \frac{n_\gamma(\epsilon) \epsilon  }{e^{\beta \epsilon }-1}d\epsilon\, ,
\end{eqnarray}
are  the energies of the electrons/positrons and photons that compose the plasma; $n_\pm(\epsilon)$ / $n(\epsilon)$ are the corresponding densities of states and $s$ are the two spin states ($s=\pm 1/2$ for the leptons and $s=\pm1$ for the photons) .  We include photons to the plasma for consistency as there might occur  pair production/annihilation in the plasma.  Furthermore, the numbers of electrons and positrons inside the cavity  are :
\begin{equation}
N_\pm=\sum_s   \int \frac{n_\pm(\epsilon)  d\epsilon}{e^{\beta (\epsilon \mp \mu )}+1} \, .
\end{equation}

All thermodynamical quantities can be obtained from the  free functions :
\begin{eqnarray}
\beta F_\pm& = - &\sum_s\int n_\pm(\epsilon)  \log \left[1+e^{-\beta (\epsilon \mp \mu )}\right]d\epsilon \, ,\\
\beta F_\gamma &= &\sum_s \int n_\gamma(\epsilon)  \log\left[1- e^{-\beta \epsilon}\right] d\epsilon \, ,
\label{F}
\end{eqnarray}
where $s$ represents the particle's spin. Indeed, we can rewrite the  conservation equations (eqs. (\ref{conservation})) as
\begin{eqnarray}
{\cal E}&=&M+\frac{\partial \beta F }{\partial \beta}  \label{Econservation} \, ,\\
{\cal Q}&=& Q+e \frac{\partial  F}{\partial \mu} \label{Qconservation}\, ,
\end{eqnarray}
where $F=F_++F_-+F_\gamma$ is the total free function of the system. The total entropy can also be obtained from the free function in the usual way:
\begin{equation}
S=\beta^2 \frac{\partial F}{\partial \beta}\, .
\end{equation}
 Integrating by parts the free energies 
\begin{eqnarray}
F_\pm& = &-\sum_s\int   \frac{\Omega_\pm(\epsilon)d\epsilon}{e^{\beta (\epsilon \mp \mu )}+1} \, ,\\
F_\gamma &= & -\sum_s \int  \frac{\Omega_\gamma(\epsilon)d\epsilon}{ e^{\beta \epsilon}-1} \, .
\label{F}
\end{eqnarray}

We proceed by counting  the total number of states up to energy $E$ for electrons/positrons and photons  $\Omega_\pm(\epsilon)$ and $\Omega_\gamma(\epsilon)$ ;  $d\Omega_\pm/d\epsilon=n_\pm$ and $d\Omega_\gamma/d\epsilon=n_\gamma$. The effective wave equation for a particle of charge $e$, spin $s$ and mass $\tilde{m}$ is given by\cite{chandra},\cite{teukolski}:
\begin{eqnarray}
&-&\frac{1}{g(r)}(\partial_t-i e A_0)^2 \psi-\frac{2s}{r}\left(1-\frac{r g'(r)}{2g(r)}\right)(\partial_t-i e A_0) \psi
+\frac{(r^2 g(r))^{-s}}{r^2} \partial_r \left((r^2 g(r))^{s+1}\partial_r \right)\psi\\
&+&\frac{1}{r^2} \left(\frac{1}{\sin\theta}\partial_\theta (\sin\theta\partial_\theta)+\frac{1}{\sin^2\theta}\partial_\varphi^2\right)\psi
+\frac{2i s}{r^2} \frac{\cos\theta}{\sin^2\theta}\partial_\varphi\psi +\frac{s-s^2 \cot^2\theta}{r^2} \psi -\tilde{m}^2\psi=0\, .
\end{eqnarray}
Shall we write
\begin{equation}
\psi=A(r,\theta) e^{i(\epsilon t-m\varphi-T(\theta)-R(r))}\, ,
\end{equation}
 Then, real  part of the wave equation equation is
\begin{equation}
\left[\frac{1}{g(r)}(\epsilon_t- e A_0)^2 -g(r) p_r^2 -\frac{p_\theta^2}{r^2}-\frac{(m-s \cos\theta)^2}{r^2 \sin^2\theta}+\frac{s}{r^2} -\tilde{m}^2\right]A+\frac{1}{r^2} \partial_\theta^2 A+(s+1)\frac{(r^2 g(r))'}{r^2} \partial_r A+g(r) \partial_r^2 A=0  \, ,
\label{real}
\end{equation}
while the  imaginary part is 
\begin{equation}
\left[\frac{2s}{r}\left(1-\frac{r g'(r)}{2 g(r)}\right) (\epsilon -eA_0) +\frac{\cot \theta\, p_\theta}{r^2}+g (\partial_r p_r)+\frac{\partial_\theta p_\theta}{r^2} +\frac{(s+1)(r^2 g(r))'}{r^2} p_r\right]A+2\frac{p_\theta \partial_\theta A+2 g(r)p_r \partial_rA } {r^2}=0 \, .
\end{equation}

In the semi-classical approximation the total number of states is the volume in phase space \cite{thooft}:
\begin{equation}
\Omega_{+s}(\epsilon)=\frac{1}{(2\pi\hbar)^3}  \sum_m \int d\varphi d\theta dr p_\theta p_r
\label{volume}
\end{equation} 
At this approximation, we consider  $A(r,\theta)$ to be a slowly varying function in eq.(\ref{real}). Performing the immediate integration in $p_\theta$ and $\varphi$ in the previous equation, isolating $p_\theta(r,\theta)$ in eq.(\ref{real}) and substituting into eq.( \ref{volume} ) it follows that  \cite{ghosh,minho}
\begin{equation}
\Omega_{+s}(\epsilon)= \frac{1}{4\pi^2 \hbar^3}\sum_m \int \left[\frac{r^2(\epsilon-eA_0)^2}{g(r)}- g(r)r^2  p_r^2 +s -\frac{(m-s\cos\theta)^2}{\sin^2\theta}
-\tilde{m}^2r^2 \right]^{1/2} dp_r d\theta dr \, .
\end{equation}
Integrating over $p_r$ in the classical allowed region (where the integrand is real) gives
\begin{equation}
\Omega_{+s}(\epsilon)= \frac{1}{8 \pi \hbar^3}\sum_m \int \left[\frac{r^2(\epsilon-eA_0)^2}{g(r)} +s -\frac{(m-s\cos\theta)^2}{\sin^2\theta}-\tilde{m}^2r^2 \right]  \frac{d\theta dr}{r\sqrt{g(r)} }\, .
\end{equation}
At this point, we convert the sum over $m$ into an integral over the region where the integrand (number of states per "unit volume") is positive and then integrate over $\theta$. The integral diverges at the horizon $r_+$ and we insert a cutoff (or either a brick wall \cite{thooft}) at the lower end of the integral:
\begin{equation}
\Omega_{+s}(\epsilon)= \frac{1}{3 \pi \hbar^3} \int_{r_++\delta}^R \left[ (\epsilon-\frac{eQ}{r})^2+\frac{s g(r)}{r^2}
-g(r)\tilde{m}^2 \right]^{3/2}  \frac{ r^2 dr}{g^2(r)} \, .
\label{omega}
\end{equation}
Performing the integral and keeping only the divergent terms at the lower end of the integral:
\begin{eqnarray}
& &\Omega_{+s}(\epsilon)= \frac{1}{3 \pi \hbar^3} \frac{r_+^6}{\delta (r_+-r_-)^2}  \left(\epsilon-\frac{eQ}{r_+}\right)^3\\
 &-&\frac{1}{3 \pi \hbar^3}\left[\left(\epsilon-\frac{e Q}{r_+}\right)^3\left(\frac{6r_+^5}{(r_+-r_-)^2}-\frac{2r_+^6}{(r_+-r_-)^3}\right) 
+  \left(\epsilon-3\frac{eQ}{r_+}\right)^2  \frac{e Q r_+^4}{(r_+-r_-)^2} +\frac{3}{2}  \left(\epsilon-\frac{eQ}{r_+}\right) \left(s-\frac{\tilde{m}}{r_+^2} \right) \frac{r_+^4}{(r_+-r_-)}
\right] \ln\left(\frac{\delta}{M}\right) \, . \nonumber \\
\end{eqnarray}
Notice that as the particle's mass $\tilde{m}$  and spin  $s$  shows up only in the logarithmic  divergence, in the leading order divergence the particle's mass and spin are immaterial .Accordingly 
\begin{equation}
F_+= -\frac{2}{3 \pi \hbar^3} \frac{r_+^6}{\delta (r_+-r_-)^2} \int_0^\infty \frac{f_+(\epsilon) d\epsilon}{e^{\beta(\epsilon-\mu)}+1} \, ,
\end{equation}
where 
\begin{equation}
f_+(\epsilon)= \left(\epsilon-\frac{e Q}{r_+}\right)^3 \, .
\end{equation}
Shifting integration variable ($\epsilon\rightarrow \epsilon +\mu$)  and then reflecting  ($\epsilon \rightarrow -\epsilon$) when required gives:
\begin{equation}
F_+= -\frac{2}{3 \pi \hbar^3} \frac{r_+^6}{\delta (r_+-r_-)^2}\left[ \int_0^\mu \frac{f_+(\mu-\epsilon)d\epsilon}{1+e^{-\beta \epsilon}} +\int_0^\infty \frac{f_+(\epsilon+\mu) d\epsilon}{e^{\beta(\epsilon)}+1}\right] \, .
\end{equation}
Similarly, shifting integration variable ($\epsilon\rightarrow \epsilon -\mu$):
\begin{equation}
F_-= -\frac{2}{3 \pi \hbar^3} \frac{r_+^6}{\delta (r_+-r_-)^2}\left[ -\int_0^\mu \frac{f_-(\epsilon-\mu)d\epsilon}{1+e^{\beta \epsilon}} +\int_0^\infty \frac{f_-(\epsilon-\mu) d\epsilon}{e^{\beta(\epsilon)}+1}\right] \, ,
\end{equation}
with
\begin{equation}
f_-(\epsilon)= \left(\epsilon+\frac{e Q}{r_+}\right)^3 \, .
\end{equation}
Since
\begin{eqnarray}
-f_-( \epsilon-\mu)=f_+(\mu -\epsilon)&=& \left(\mu -\epsilon-\frac{e Q}{r_+}\right)^3 \, ,\\
f_+( \epsilon+\mu)+f_-(\epsilon-\mu)&=&2\epsilon^3 +6\epsilon \left(\mu-\frac{eQ}{r_+}\right)^2 \, ,
\end{eqnarray}
then,
\begin{equation}
F_++F_-= -\frac{2}{3 \pi \hbar^3} \frac{r_+^6}{\delta (r_+-r_-)^2}\left[ \int_0^\mu  \left(\mu -\epsilon-\frac{e Q}{r_+}\right)^3d\epsilon  +2 \int_0^\infty \frac{\epsilon^3 d\epsilon}{e^{\beta(\epsilon)}+1}+6\left(\mu-\frac{eQ}{r_+}\right)^2
 \int_0^\infty \frac{\epsilon d\epsilon}{e^{\beta(\epsilon)}+1}
\right]\, .
\end{equation}\
At last,  the total free energy of the plasma is:
\begin{equation}
F= -\frac{8}{3 \pi \hbar^3\Delta^2} \frac{r_+^8}{ (r_+-r_-)^3}\left[  \frac{1}{4} \left(\mu -\frac{e Q}{r_+}\right)^4-\frac{1}{4} \left(\frac{e Q}{r_+}\right)^4+\frac{11\pi^4}{60 \beta^4} +\frac{\pi^2}{2\beta^2} \left(\mu-\frac{eQ}{r_+}\right)^2\right] \, ,
\end{equation}
where 
\begin{equation}
\Delta=\int_{r_+}^{r_++\delta}\sqrt{g_{rr}} dr= \frac{r}{\sqrt{(r-r_+)(r-r_-)}}\approx \frac{2 r_+ \sqrt{\delta}}{\sqrt{r_+-r_-}}
\end{equation}
is the proper length from the horizon.
The system's  entropy reads
\begin{equation}
S= \frac{\pi r_+^2}{\hbar}+\frac{8}{3 \pi \hbar^3} \frac{r_+^8}{\Delta^2 (r_+-r_-)^3}\left[  +\frac{11\pi^4}{15 \beta^3} +\frac{\pi^2}{\beta} \left(\mu-\frac{eQ}{r_+}\right)^2\right] \, ,
\end{equation}
where the first term corresponds to the Bekenstein-Hawking entropy. The first derivative of this entropy with respect to the mass, while enforcing the the energy conservation constraint yields the black hole temperature:
\begin{equation}
\beta^{-1}=\frac{\hbar(r_+-r_-)}{4\pi r_+^2}\, 
\end{equation}
The second derivative of the entropy with respect to the mass gives the stability regime, which was studied extensively in the past \cite{hawking}-\cite{pavon2}. 
It is convenient to define the dimensionless parameter
\begin{equation}
\alpha=\frac{\beta \left(\mu-\frac{eQ}{r_+}\right) }{\pi}\, .
\end{equation}
 Then we can express 
\begin{equation}
S= \frac{\pi r_+^2}{\hbar}+  \frac{11 r_+^2}{360 \Delta^2 } +  \frac{r_+^2}{24 \Delta^2}\alpha^2 \, .
\end{equation}
Assuming the system plasma+ BH is neutral,  the charge conservation \ref{Qconservation} condition reads:
\begin{equation}
 \alpha^3+\alpha+\frac{24 Q\pi }{e} \frac{\Delta^2 }{r_+^2} =0 \, .
\end{equation}
At the lowest order in $Q$ we approximate $r_+^2=4M^2 -2 Q^2$ and $\alpha = -(6 \pi \Delta^2 Q)/e M^2 $. Accordingly, 
\begin{equation}
S\approx  \left[\frac{4\pi M^2 }{\hbar}+  \frac{11 M^2 }{90 \Delta^2 }\right] - \left[\frac{2\pi  }{\hbar}  +\frac{11   }{180 \Delta^2 } -
 \frac{24 \pi^2 \Delta^2}{e^2 M^2}\right] Q^2 \, .
\end{equation}
Since $\Delta<< \sqrt{e M}$, we can neglect the last term in the second squared bracket. Notice $(\partial S/\partial Q)_{Q=0}$  vanishes identically.  

Recalling that fluctuations are given by \cite{landau}:
\begin{equation}
\frac{1}{\Delta Q^2}=- \frac{\partial^2 S}{\partial Q^2 }\, ,
\end{equation}
then the lengthy calculation boils down to a very simple result:
\begin{equation}
\frac{1}{\Delta Q^2}= \frac{1}{\Delta Q^2_{BH}}+ \frac{1}{\Delta Q^2_{plasma}} \, ,
\end{equation}
with
\begin{equation}
\Delta Q_{BH} =\sqrt{\frac{\hbar}{4\pi}} \quad ;  \quad \Delta Q_{plasma}=\sqrt{\frac{90 }{11}}\Delta \,. 
\end{equation}

\section{Concluding remarks}

Since the density of modes becomes very large at the horizon, we have seen that the leading contribution to the thermodynamical quantities is of order $\delta^{-1}$, where $\delta$ is the distance from the horizon (or, equivalently the width of the brick wall) . In this case, either the field's mass or the spin  are immaterial . Furthermore, we. have seen that the contribution of $\alpha$ to the free energy can be also neglected. Then, inspecting eq.(\ref{omega}) we can write 
\begin{equation}
F_{fields}\sim -\frac{1}{\hbar^3  \beta^4} \int \frac{r^2 dr}{g^2(r)}\, ,
\end{equation}
where we dropped numerical factors.  Consequently, the total energy and entropy are
\begin{eqnarray}
E_{fields}& \sim& \frac{1}{\hbar^3} \int T_{local}^4 r^2 dr\\
S_{fields} &\sim& \frac{1}{\hbar^3} \int T_{local}^3 \sqrt{g_{rr}}r^2 dr
\end{eqnarray}
with $T_{local}= 1/\beta \sqrt{g(r)}$ is Tolman's temperature ,  the one  measured by a local observer at rest \cite{tolman,visser}.  The energy density $\rho=a T_{local}^4$ embodies also the gravitational (self) energy which is $\rho(1-\sqrt{g_{rr}})$ \cite{MTW}. 
The   field's  fluctuations  happen  within  a thin envelope, the quantum atmosphere of the black hole.  Although we were  careful enough to take into account charge conservation,  surprisingly the field's contribution  $\Delta Q_{plasma}$ does not depend upon  the charge of the charge carriers! Furthermore,  should we take the cut-off parameter to be Planckian $\Delta \sim \sqrt{\hbar}$, the two contributions $\Delta Q_{plasma}$ and $\Delta Q_{BH}$  differ only by  a numerical factor of order one. If the two contributions to the entropy are the same order, was it appropriate to neglect the field's back reaction to the geometry? Recalling  that the dominant contribution to he energy within the thin envelope (the  quantum atmosphere )
\begin{equation}
E\sim \left(\frac{r_+-r_-}{r_+}\right)^2  \frac{\hbar}{\delta} = 4\pi r_+^2 \rho \delta
\end{equation}
Equivalently, the total energy density is 
\begin{equation}
\rho \sim \frac{(r_+-r_-)^2}{r_+^4}  \frac{\hbar}{\delta^2}
\end{equation}
From Einstein's equation
\begin{equation}
\left(r(1-\frac{1}{g_{rr}})\right)'=8\pi r^2 \rho
\end{equation}
Thus assuming that all energy is concentrated across the thin shell of width $\delta$ around the horizon $r_+$ the discontinuity it produces in $1/g_{rr}$ is
\begin{equation}
D \frac{1}{g_{rr}}\sim -\frac{(r_+-r_-)^2}{r_+^3}  \frac{\hbar}{\delta} \sim -\left(\frac{r_+-r_-}{r_+}  \right)\frac{\hbar}{\Delta^2 } \sim-  \frac{\hbar}{\Delta^2}
\end{equation}
Thus, if the width of the quantum atmosphere is, say, one order of magnitude larger than Planck's length, then the back reaction to the geometry can be neglected at a first approximation. Otherwise, if the width is much smaller than Planck's length, then the energy density produces a noticeable contribution to  the effective mass at infinity , an interpolating metric within the quantum atmosphere must be considered. Back reaction could change considerably the present discussion. 

Assuming a Planckian scale for the width of the quantum atmosphere, the field contribution to the charge fluctuation is nearly identical to that of the Bekenstein-Hawking , both being independent  upon the charge of the charge carrier. Accordingly it is legitimate to regard the quantum atmosphere as an integral part of the black-hole itself.

Should the fluctuations originate mainly as the fluctuation of a single unit of  charge $\pm  q $ with $
\overline{Q}=0$ and  $\Delta Q^2 = q^2/2$ then  $q\sim \sqrt{\hbar} $, which is not very far from the electron charge . The physical origin of the fluctuations is elusive. As  $\Delta Q_{plasma}$ does not depend upon the charge of the charge carries, the black hole charge fluctuation cannot be regarded as resulting from the  trade of charges between the BH and the plasma. Should we regard that charge is trapped inside the black hole and  fluctuations  arise due an ensemble of identical BH's ? If not, could the culprit for the fluctuations be  charges crossing the Cauchy horizon $\overline{r_-} \approx \Delta Q^2/2M \sim \hbar/M$ in both directions? Since the origin of fluctuations is the Bekeinstein-Hawking entropy , the origin of this paradox is deeply rooted into the semi-classical quantum gravity . 
\section*{aknowledgments} I am thankful to Mikhail Zubkov for his criticism and comments

\end{widetext}
\end{document}